\documentclass[paper]{JHEP3}

\usepackage{epsfig,bbm}

\newcommand{\deldag}{\mathbin{\partial\mkern-10.5mu\big/}}

\newcommand{\be}{\begin{equation}} 
\newcommand{\ee}{\end{equation}}
\newcommand{\bea}{\begin{eqnarray}} 
\newcommand{\eea}{\end{eqnarray}}

\newcommand{\bmp}{\noindent\begin{minipage}{16cm}}
\newcommand{\emp}{\end{minipage}\vskip 7mm} 
\def\lsim{\mathrel{\raise.3ex\hbox{$<$\kern-.75em\lower1ex\hbox{$\sim$}}}}
\def\gsim{\mathrel{\raise.3ex\hbox{$>$\kern-.75em\lower1ex\hbox{$\sim$}}}}
\newcommand{\ksl}{\mathbin{k\mkern-10mu\big/}}
\newcommand{\psl}{\mathbin{p\mkern-10mu\big/}}
\newcommand{\intron}[1]{}

\def\sfrac#1#2{{\textstyle\frac{#1}{#2}}}

\title{Naturality, unification and dark matter}

\author{Kimmo Kainulainen\footnote{kimmo.kainulainen@jyu.fi}\\
        \\ Department of Physics, P.O.Box 35 (YFL), 
        \\ FI-40014 University of Jyv\"askyl\"a, Finland, 
        \\ and 
  	    \\ Helsinki Institute of Physics, P.O.~Box 64, 
  	    \\ FI-00014 University of Helsinki, Finland.\\}
\author{Kimmo Tuominen \footnote{kimmo.tuominen@jyu.fi}\,\,\footnote{On leave of absence from Department of physics, University of Jyv\"askyl\"a} \\
{CP}$^{ \bf 3}${-Origins}, 
Campusvej 55,\\ DK-5230 Odense M, Denmark \\and\\
Helsinki Institute of Physics, 
P.O.Box 64, \\ FI-000140, University of Helsinki, Finland}
\author{Jussi Virkaj\"arvi \footnote{jussi.virkajarvi@jyu.fi}
        \\
        \\ Department of Physics, P.O.Box 35 (YFL), 
        \\ FI-40014 University of Jyv\"askyl\"a, Finland, 
        \\ and 
  	    \\ Helsinki Institute of Physics, P.O.~Box 64, 
  	    \\ FI-00014 University of Helsinki, Finland.\\}

\abstract{We consider a model where electroweak symmetry breaking is driven by Technicolor dynamics with minimal particle content required for walking coupling and saturation of global anomalies. Furthermore, the model features three additional Weyl fermions singlet under Technicolor interactions, which provide for a one-loop unification of the Standard Model gauge couplings. Among these extra matter fields exists a possible candidate for weakly interacting dark matter. We evaluate the relic densities and find that they are sufficient to explain the cosmological observations and avoid the experimental limits from earth-based searches. Hence, we establish a non-supersymmetric framework where hierarchy and naturality problems are solved, coupling constant unification is achieved and a plausible dark matter candidate exists.}


\preprint{CP3-Origins-2010-3}

\begin{document}

\section{Introduction}

There is a large amount of evidence that dark matter (DM) is abundant in the universe from the galactic to the Hubble horizon scale, but the true nature of the DM remains a mystery. Although a great number of particle physics motivated dark matter candidates have been proposed over the years, we still miss a direct observation of the DM and the models have to be, at least in part, measured on the basis of their theoretical appeal. By far the most popular DM candidate in this respect has been the lightest supersymmetric particle, LSP, which naturally arises in the supersymmetric extensions of the standard model. Indeed, the most remarkable features of the supersymmetry are the solution to the hierarchy problem, possibility of a gauge coupling constant unification and the existence of dark matter particle. In this paper we point out that all these features can be provided in the context of recent walking Technicolor (TC) theories without a need for supersymmetry. 

Fundamental non-supersymmetric scalar fields require fine-tuning and are unnatural. This behavior is also known as the hierarchy problem. When electroweak symmetry breaking arises dynamically, as in Technicolor theories \cite{Sannino:2008ha,Hill:2002ap}, only fermions and gauge fields appear as fundamental constituents and the scalar degrees of freedom are composite bound states under the new Technicolor interaction; the hierarchy problem is thus trivially solved under TC. The most severe phenomenological constraints guiding Technicolor model building are the suppression of flavor changing neutral current interactions (when fermion mass generation due to extended Technicolor interactions \cite{Dimopoulos:1979es,Eichten:1979ah,Appelquist:2003hn} or the like \cite{Simmons:1988fu,Dine:1990jd,Samuel:1990dq, Kagan:1990az,Carone:1992rh,Antola:2009wq} is accounted for) and the smallness of the contributions to the precision $S$-parameter from beyond Standard model (SM) matter fields. The FCNC constraint seems to imply that the evolution of the coupling constant of the non-Abelian gauge theory underlying the TC sector should be governed by a quasi-stable infrared fixed point, {\em i.e.}~the coupling should ``walk" rather than run over a large hierarchy of scales. The precision constraints then require that this walking behavior should be obtained with a modest number of new matter fields. Originally in~\cite{Sannino:2004qp} and further in~\cite{Dietrich:2005jn}, it was suggested that simple consistent models can be built by considering technifermions transforming under higher representations of the gauge group. In particular the choice of SU(2) gauge group with two adjoint Dirac flavors was identified as the minimal model with the smallest possible number of new degrees of freedom. This model, termed Minimal Walking Technicolor (MWT), has recently been studied nonperturbatively through lattice simulations~\cite{Catterall:2007yx,DelDebbio:2008wb,Hietanen:2008mr}, with results supporting the original conjecture that the model is infrared conformal for massless techniquarks. The collider phenomenology of this model has also been studied~\cite{Foadi:2007ue,Foadi:2007se,Antipin:2009ks,Frandsen:2009fs}.

However, the requirement of naturality is but one possible paradigm for the  model building and others, like unification of the three SM coupling constants, can be considered. The Minimal Supersymmetric Standard Model (MSSM) provides this behavior, lending strong appeal for supersymmetry. However, as noted in the literature~\cite{Li:2003zh,Gudnason:2006mk}, only a very small portion of the degrees of freedom in the supersymmetric spectrum are actually needed to achieve this unification. One might then ask also what is the minimal extra matter content required for a successful unification of the SM coupling constants in a given Technicolor model for electroweak symmetry breaking. This question was answered in~\cite{Gudnason:2006mk}, where it was shown that in the context of the MWT model, unification can be achieved by adding one Weyl fermion in the adjoint representation of QCD color and one Weyl fermion in the adjoint representation of the weak SU$_L(2)$. In terms of their quantum numbers these fields are similar to the gluino and wino of the MSSM. 

The third main paradigm of model building comes from cosmology: a stable weakly interacting massive particle is the favored candidate for dark matter with relic density of thermal origin\footnote{Recently there has been interest in {\em{asymmetric}} type DM candidates, see 
e.g.~\cite{Ryttov:2008xe,Foadi:2008qv,Frandsen:2009mi}.}, and it would be desirable to have a particle physics model set-up where such a particle arises naturally. Again, in the MSSM such a candidate is provided by the LSP. Technicolor models also feature natural dark matter candidates and in particular the lightest electrically neutral technibaryon is an appealing possibility~\cite{Gudnason:2006yj}. The WIMPs suggested in the context of MWT model include exotics \cite{Kouvaris:2007iq} and a fourth generation neutrino~\cite{Kainulainen:2006wq}. The latter scenario was extended to the case of a fourth generation neutrino mixing with sterile state in ref.~\cite{Kainulainen:2009rb} (for an early related work see refs.~\cite{Enqvist:1990yz,Enqvist:1988dt}). Also strongly interacting and composite MWT-motivated DM particles have been considered~\cite{Khlopov:2007ic}.
The purpose of this paper is to investigate the sub-weakly interacting dark matter candidate in the context of the unified Technicolor model described above. The new WIMP is identified as a mixture of the neutral member of the new adjoint SU$_L$(2) triplet needed for the coupling constant unification and an additional singlet Weyl fermion (corresponding to the bino of the MSSM). We show that this scenario is only weakly constrained by the current observational data and that the new WIMP can be dark matter if its mass is $m_2 \gsim 70$ GeV. Finally, for  $m_2 \approx 80$ GeV (and with a light composite Higgs boson mass $m_H \approx 200$ GeV) the model would be consistent with the recent CDMS II collaboration~\cite{Ahmed:2009zw} dark matter signal.

This paper is organized as follows: In section \ref{model} we describe the basic building blocks of the model concentrating especially on the weak currents, the unification of gauge couplings, the mass generating mechanisms and mass mixing scheme which gives rise to the new WIMP and its interactions necessary for the evaluation of the DM relic density. In section \ref{results} we present the results of our numerical analysis for the relic abundances and summarize the current and some of the projected future observational constraints on the model parameters. In Sec. \ref{outlook} we conclude and briefly outline the prospects of future studies within this model, and finally in the Appendix we give some details of the cross section calculations.
  
\section{The model}
\label{model}

We consider the MWT extension of the SM. This means that we replace the Higgs sector with a new SU(2) gauge theory with two Dirac flavors in the adjoint representation of the gauge group. The two flavors, $U$ and $D$, are arranged into a doublet of SU$_L(2)$, and taking the Technicolor degree of freedom into account implies that we are adding three doublets of SU$_L(2)$ and the resulting theory is anomalous. To cure this anomaly, one further weak doublet $L$ carrying no QCD or Technicolor quantum numbers is introduced. The gauge anomalies are cancelled by the following hypercharge assigments
\bea
Y(U_L) &=& Y(D_L)=\frac{1}{2}y,~~Y(U_R)=\frac{1}{2}(y+1),~~Y(D_R)=-\frac{1}{2}(y-1),\nonumber \\
Y(E_L) &=& Y(N_L)=-\frac{3}{2}y,~~Y(N_R)=\frac{1}{2}(-3y+1),~~Y(E_R)=-\frac{1}{2}(-3y-1),
\eea
where $y$ is any real number. We will choose $y=1/3$ throughout this paper. While other choices are possible and phenomenologically viable, this particular choice is interesting since it makes the techniquarks and fourth generation leptons appear as on ordinary SM family from the electroweak interaction viewpoint.
 
As discussed in the introduction, it has been shown~\cite{Gudnason:2006mk} that  one may achieve a good one loop unification of the SM coupling constants in the MWT model by adding only a modest number of Weyl fermions to MWT. The new fields have quantum numbers similar to the gauginos of MSSM: one Weyl fermion in the adjoint representation of QCD color and one Weyl fermion in the adjoint representation of SU$_L(2)$.  Here we consider the situation where the ``gluino" (the SU$(3)$ adjoint field) is heavy and decoupled from the low-energy theory. However, we will introduce an additional light singlet Weyl fermion corresponding to the bino of the MSSM. Guided by these arguments we consider the following addition to the low-energy Lagrangian of the standard model:
\begin{equation}
{\mathcal{L}} =  i\omega^{\dagger}\bar{\sigma}^\mu D_\mu \omega 
               + i\beta\sigma^\mu\partial_\mu \beta^{\dagger} + \cdots \,,
\label{eq:Lstart}
\end{equation}
where $\omega = (w^1,w^3,w^3)$ is the left-handed triplet and $\beta^{\dagger}$ the right-handed singlet. $D_\mu$ is the covariant derivative, which in the component form reads
\begin{equation}
D^{ac}_\mu =  \partial_\mu \delta^{ac} + g \epsilon^{abc} A^b_\mu \,.
\label{eq:covder}
\end{equation}
Here we used $[T^a]_{bc}= -i\epsilon^{abc}$ for the SU$_L$(2) generators in the adjoint representation. That is, the field $\omega$ transforms as a triplet under SU$_L(2)$ and is a singlet under hypercharge. The field $\beta^\dagger$ is singlet under all SM gauge groups. The dots in Eq.~(\ref{eq:Lstart}) refer to the couplings with the effective scalar fields which give rise to masses; they will be introduced explicitly in Sect.~\ref{sec:massterms}.

\subsection{Weak currents}

We now wish to rewrite the lagrangian Eq.~(\ref{eq:Lstart}) in the usual four-component notation, in terms of charge eigenstate Dirac- and Majorana fields. We start by defining the eigenstates of the diagonal operator $T^3$:
\begin{eqnarray}
w^\pm=\frac{1}{\sqrt{2}}(w^1\mp iw^2),~~ w^0  =w^3
\end{eqnarray}
which have electromagnetic charge $\pm 1$ and 0, as the notation suggests, due to $Q=T_3+Y$. Using these fields one finds, for example,
\begin{equation}
i \omega^{\dagger}\bar{\sigma}^\mu\partial_\mu \omega = 
 iw^{+\dagger}\bar{\sigma}^\mu\partial_\mu w^+
+iw^{-\dagger}\bar{\sigma}^\mu\partial_\mu w^-
+iw^{0\dagger}\bar{\sigma}^\mu\partial_\mu w^0 \,.
\label{free_L}
\end{equation} 
Similarly the gauge interaction term
\begin{equation}
{\mathcal{L}}_{\rm gauge} = ig \epsilon^{abc} w^{a\dagger} \bar{\sigma}^\mu W_\mu^b w^c \,
\label{eq:gaugeint1}
\end{equation}
can be easily rewritten in terms of $w^{\pm}$ and the charged gauge bosons $W_\mu^\pm = (W^1_\mu \mp i W^2_\mu)/\sqrt{2}$. All expressions can then be transformed from the Weyl notation to the 4-dimensional Dirac notation by defining Dirac spinors carrying negative and positive charge:
\begin{equation}
w_D^{-}=\left(\begin{array}{c} {w}^-_\alpha \\ (w^+)^{\dagger\dot{\alpha}}\end{array}\right),~~
w_D^{+}=\left(\begin{array}{c} {w}^+_\alpha \\ (w^-)^{\dagger\dot{\alpha}}\end{array}\right)
\end{equation}
and neutral 4-component Majorana spinors
\begin{equation}
w_M^0=\left(\begin{array}{c} {w}^0_\alpha \\ 
(w^0)^{\dagger \dot{\alpha}}\end{array}\right), ~~
\beta_M=\left(\begin{array}{c} \beta_\alpha \\ 
\beta^{\dagger\dot{\alpha}}\end{array}\right) \,.
\label{eq:neutral}
\end{equation}
These satisfy $(w_D^{-})^{c}=w_D^+$, $(w_M^{0})^{c}=w_M^0$ and $(\beta_M)^{c}=\beta_M$.  Since only one of the charged spinors can be treated as independent degree of freedom, we can write everything either using only $w_M^0$, $\beta_M$  and either of the charged Dirac-spinors $w_D^\pm$. Setting $w_D\equiv w_D^-$ we can rewrite the Lagrangian (\ref{eq:Lstart}) as
\begin{eqnarray}
{\mathcal{L}} 
&=& i\bar w_D \deldag w_D 
  + i\bar w_M \deldag w_M 
  + i\bar \beta_M \deldag\beta_M  
\nonumber \\
&+& g\left(
    W_\mu^+\overline{w}^0_M\gamma^\mu w_D
   +W_\mu^-\overline{w}_D\gamma^\mu w^0_M
   -W^3_\mu\overline{w}_D\gamma^\mu w_D \right) + \cdots \,,
\label{gauge_lagrangian}
\end{eqnarray}
The field $\beta_M$ does not couple to electroweak gauge fields but may mix with the neutral wino through the mass terms, again represented by dots in Eq.~(\ref{gauge_lagrangian}). We can immediately read the charged and neutral currents from (\ref{gauge_lagrangian}):
\begin{eqnarray}
\label{ew_currents}
{\mathcal{L}}_W &=& g(W_\mu^+\overline{w}^0_M\gamma^\mu w_D+W_\mu^-\overline{w}_D\gamma^\mu w^0_M),\\
{\mathcal{L}}_Z &=& g\cos\theta_W Z_{\mu}\overline{w}_D\gamma^\mu w_D, \nonumber \\
{\mathcal{L}}_A &=& e A_{\mu}\overline{w}_D\gamma^\mu w_D.\nonumber
\end{eqnarray}
It is very important to note that the field $w_M^0$ does not couple to the neutral gauge boson $Z_\mu$. Therefore, we will eventually need only the charged current in our dark matter analysis.

\subsection{Unification}

Let us now briefly review how the model described above leads to a good one-loop unification of the SM coupling constants. Generally, the evolution of the coupling constant $\alpha_n$ of an SU$(n)$ gauge theory at one loop level is controlled by
\begin{equation}\label{running}
{\alpha_{n}^{-1}(\mu) = \alpha_{n}^{-1}(M_Z) - \frac{b_n}{2\pi}\ln
\left(\frac{\mu}{M_Z}\right) \ .}
\end{equation}
For SM we have three coupling constants corresponding to  SU$(3)\times$SU$(2)\times$U$(1)$ for $n=3,2,1$. In the above equation the first coefficient $b_n$ of the beta function is 
\begin{equation}
 b_n = \frac{2}{3}T(R) N_{wf} + \frac{1}{3} T(R')N_{cb} -
 \frac{11}{3}C_2(G) \ .
\label{eq:bn}
\end{equation}
where $T(R)$ is the Casimir of the representation $R$ to which the
$N_{wf}$ Weyl fermions belong and $T(R')$ is the Casimir of the representation $R'$
to which the $N_{cb}$ complex scalar bosons belong.  Finally, $C_2(G)$ is the
quadratic Casimir of the adjoint representation of the gauge group.

Now we require the SM coupling constants to unify at some very high energy scale $M_{GUT}$. This means that the three couplings are all equal at the scale $M_{GUT}$, i.e. $\alpha_3(M_{GUT})= \alpha_2(M_{GUT})=\alpha_1(M_{GUT})$ with $\alpha_1= \alpha/(c^2\cos^2 \theta_W)$ and $\alpha_2 = \alpha/ \sin^2\theta_W$, where $c$ is a normalization constant to be determined by the choice of the unifying group (like the paradigmatic SU(5)).
Assuming one-loop unification using Eq.~(\ref{running}) for $n=1,2,3,$ one
finds the following relation 
\begin{equation} 
\label{unification}
B \equiv
\frac{b_3-b_2}{b_2-b_1} = 
      \frac{\alpha_3^{-1} -\alpha^{-1}\sin^2\theta_W}
           {(1+c^2)\alpha^{-1} \sin^2\theta_W-c^2\alpha^{-1}} \ .
\end{equation}
In the above expressions the Weinberg angle $\theta_W$, the electromagnetic coupling constant $\alpha$ and the strong coupling constant $\alpha_3$ are all evaluated at the scale $\mu=M_Z$. For a given particle content, we denote the LHS of Eq.~(\ref{unification}) by $B_{\rm theory}$ and the RHS by $B_{\rm exp}$. Whether $B_{\rm theory}$ and $B_{\rm exp}$ agree is a simple way to check if the coupling constants unify. However, finding a convergence of all coupling constants at a common scale is not enough; to have the proton decay under control the unification scale has to be sufficiently large. 
With one-loop running the unification scale is given by the expression
\begin{equation}
M_{\rm GUT} = M_Z \exp 
       \left[ {2\pi \frac{(1+c^2)\alpha^{-1}\sin^2\theta_W-c^2\alpha^{-1}}
                         {b_2-b_1}} \right]\,. 
\label{massu}
\end{equation}

To be specific, we will use the experimental values from ref.~\cite{Amsler:2008zzb}: $\sin^2 \theta_W (M_Z) = 0.23119\pm 0.00014$, $\alpha^{-1}(M_Z) = 127.909\pm 0.019$, $\alpha_3(M_Z) =0.1217\pm 0.0017$ and $M_Z = 91.1876\pm 0.0021$ GeV. We also take $c=\sqrt{3/5}$, corresponding to the SM case with $N_g$ ordinary matter generations. (This result remains valid also when the hypercharge is upgraded to one of the generators of $SU(5)$). With these numerical values we find from Eqs.~(\ref{unification}) and (\ref{massu}) 
the following conditions for a successful 1-loop unification:
\begin{eqnarray} B_{\rm theory} = 
\frac{b_3-b_2}{b_2-b_1} \approx 0.725 \quad {\rm and }\quad
{M_{\rm{GUT}} \approx M_Z \exp\left[\frac{186}{b_1-b_2}\right] \gsim 10^{15}{\rm GeV}\,. }
\label{eq:bounds}
\end{eqnarray}

The key feature which motivates the particle content of the model we consider here, has been nicely explained in \cite{Li:2003zh}: At one-loop contributions to $b_3 - b_2$ or $b_2 - b_1$ emerge only from particles {\em not} forming complete representations of the unifying gauge group (like the five and the ten dimensional representations of $SU(5)$). For example the gluons, the weak gauge bosons and the Higgs particle of the SM do not form complete
representations of $SU(5)$ but ordinary quarks and leptons do.\footnote{Here we mean that these particles form complete representations of $SU(5)$, all the way from the unification scale down to the electroweak scale. The particles not forming complete representations at low energies will presumably combine at the unification scale with new particles to form complete representations of the unified gauge group. Note also, that while there is no contribution to the unification point of the particles forming complete representations, the running of each coupling constant is affected by all of the particles present at low energies.}
Let us first consider the standard model case with $N_g$ ordinary matter generations. The beta function coefficients are easily found from Eq.~(\ref{eq:bn}):
\begin{eqnarray}
b_3 & = & =\frac{4}{3}N_g -11 \nonumber\\
b_2 & = & =\frac{4}{3} N_g - \frac{22}{3} + ( \frac{1}{6} )
\nonumber\\
b_1 & = & =\frac{4}{3}N_g + ( \frac{1}{10} ).
\label{eq:bnSM}
\end{eqnarray}
Here we used $T(R)=1/2$ for the fundamental representation and appropriate $N_{wf}$ for the corresponding gauge groups. Also, for $U(1)_Y$ gauge group and given matter field, we have $T(R)=c^2Y^2=(3/5)Y^2$, where $Y$ is the corresponding hypercharge. The bracketed contributions in Eq.~(\ref{eq:bnSM}) arise from the SM Higgs field.  It is clear that the SM does not unify since $B_{\rm theory} \simeq 0.53$, independent of $N_g$, while the required value was $B_{\rm exp} \simeq 0.725$. Eq.~(\ref{eq:bnSM}) shows explicitly that only the gauge bosons and the SM Higgs are relevant for computing $B_{\rm theory}$, while the contributions from quarks and leptons drop out in the differences $b_i-b_j$, in agreement with the fact that they form complete representations of the unifying gauge group. 

Let us now turn to the model under consideration. Instead of the Higgs sector of the SM we have the MWT model with hypercharge assignment rendering the Technicolor sector identical to one extra SM generation from the electroweak interaction viewpoint as explained in the beginning of Sec.~\ref{model}. In addition we have one strongly interacting adjoint Weyl fermion which affects the running of the QCD coupling and one weak singlet affecting the running of $\alpha_2$. Using $T(G)=3,2$ for the adjoint representations of SU$(3)$ and SU$(2)$, respectively, we then find From Eq.~(\ref{eq:bn}): 
\begin{eqnarray}
b_3 & = & \frac{4}{3}N_g -11 + \frac{2}{3}\times 3 \nonumber\\
b_2 & = & \frac{4}{3} (N_g+1) - \frac{22}{3} +\frac{2}{3}\times 2 \nonumber\\
b_1 & = & \frac{4}{3}(N_g +1).
\end{eqnarray}
Using the above in (\ref{eq:bounds}), we obtain $B_{\rm{theory}}\approx 0.722$ and $M_{\rm{GUT}}\approx 3\times 10^{15}$ GeV, {\em i.e.}~an almost perfect one-loop unification and a proton lifetime well compatible with the current experimental limits.

As a final remark on unification we note that here we have only considered unification of the SM coupling constants, while in our model we have an additional gauge coupling related to the strong Technicolor interactions. One can also consider unification of all four coupling constants as has been done e.g. in \cite{Gudnason:2006mk}, but we do not pursue this here. We simply take the SM coupling constant unification as an additional motivation for the particle content which we have introduced here.

\subsection{Effective mass terms}
\label{sec:massterms}

Within the dynamical symmetry breaking framework the elementary fermion masses arise generically from extended Technicolor (ETC) interactions. Since we do not want to embark on the ETC model building, we choose to parametrize the underlying structures by considering the interactions of the fermions with the scalar sector. The strongly interacting Technicolor sector in the MWT model is, at energies of the order of the electroweak scale, best described in terms of effective theory formulated in terms of Technicolor neutral scalar (and vector) states. This effective Lagrangian has been built in \cite{Foadi:2007ue} where also the coupling to electroweak currents has been worked out as well as the couplings with SM fermions. We will use here the result that, for the hypercharge assignement we have chosen, only the SM-like part of the scalar sector couples to SM fermions. In particular we will re-introduce a single effective SM-like Higgs doublet $H$ to generate masses for fermions. The Majorana mass for the triplet is then generated by the interaction 
\begin{equation}
{\mathcal{L}}_{\omega H}
  =\frac{\lambda_L}{\Lambda} H^\dagger\omega\omega H + {\rm c.c.},
\label{doublet}
\end{equation}
where $\omega\equiv \omega^a\tau^a$ and $\tau^a=\sigma^a/2$ in terms of the Pauli matrices. The suppressing scale $\Lambda$ is related to the more complete ultraviolet theory (presumably ETC-like) expected to generate the full flavor structure emerging above the electroweak scale. For our bottom-up model building the parametrization provided by the above equation is sufficient however. When $\sqrt{2}H\rightarrow (0,v+h)^T$ Eq.~(\ref{doublet}), written in the 4-component notation, becomes
\begin{equation}
{\mathcal{L}}_{\omega H} \rightarrow
        ( M_L \overline{w}_D w_D 
   + \frac{M_L }{2}\overline{w}_M^0w_M^0 ) \,\Big(1 + \frac{h}{v} \Big)^2 \,,
\label{doublet2}
\end{equation}
where $M_L \equiv \lambda_Lv^2/4\Lambda$ simultaneously gives the mass of the charged $w_D$ field and the Majorana mass of the state $w_M^0$ constructed from the adjoint triplet. Note that the interaction Eq.~(\ref{doublet}) has a $Z_2$ symmetry corresponding to $w\rightarrow -w$.

We can construct other gauge-invariant dimension five Yukawa intercation terms using the isosinglet charge neutral field $\beta^\dagger$. In particular the mixing between the singlet and triplet fields is generated by:
\begin{eqnarray}
{\mathcal{L}}_{\omega \beta H} =
 \frac{\lambda_D}{\Lambda} \beta H^\dagger\omega H + {\rm{c.c.}} \,,
\label{mixmass}
\end{eqnarray}
which upon condensation $\sqrt{2}H\rightarrow (0,v+h)^T$ becomes:
\begin{equation}
{\mathcal{L}}_{\omega \beta H} \rightarrow
m_D (\overline w^0_L \beta_R + \overline {w}^0_R\beta_L ) \, \Big(1 + \frac{h}{v} \Big)^2 + {\rm{c.c.}}\,,
\label{mixmass2}
\end{equation}
with $m_D \equiv \lambda_Dv^2/2\Lambda$. Here we used the shorthand 4-component notations $w^0_L \equiv w^0_{ML} = (w^0_\alpha,0)^T$ and $\beta_R \equiv \beta_{MR} = (0,\beta^{\dagger\dot \alpha})^T$, and so on. The interation Eq.~(\ref{mixmass}) does not connect the adjoint fermions with ordinary matter and hence does not endanger the stability of our DM candidate. Note that it is also invariant under the extended $Z_2$ symmetry $\beta\rightarrow -\beta$ and $w\rightarrow -w$, which can be motivated by the assumption that $w^a$ and $\beta^\dagger$ transform under the same representation under possible grand unification. 

Finally we can write an interaction between the $\beta^\dagger$ and $H$ which is gauge invariant and also again respects the $Z_2$-symmetry:
\begin{eqnarray}
{\mathcal{L}}_{\beta H}&=&\frac{\lambda_R}{\Lambda} \beta \beta H^\dagger H + {\rm{c.c}}. 
\nonumber \\ 
&\rightarrow&
\frac{M_R}{2}{\overline \beta}_M\beta_M \, \Big(1 + \frac{h}{v} \Big)^2 \,,
\label{Higgs}
\end{eqnarray}
where $M_R \equiv \lambda_Rv^2/\Lambda$ is a right-handed Majorana mass for the $\beta_M$-field. In this scenario the singlet field interacts with the SM-like Higgs state. Alternatively the $\beta_M$ field can receive its mass in a dynamical symmetry breaking from the VEV of a new weak SU(2) singlet field $S$, which can plausibly emerge from a more complete extended Technicolor theory. The gauge- and $Z_2$ symmetric interaction Lagrangian for $\beta^\dagger$ and $S$ is 
\begin{eqnarray}
{\mathcal{L}}_{\beta S}= y_R S \beta\beta + {\rm{c.c.}},
\label{singlet}
\end{eqnarray}
which in symmetry breaking with $\sqrt{2}S \rightarrow v_s + s$ again results in a mass term $\frac{M_R}{2}\overline{\beta}_M\beta_M$, now with $M_R \equiv y_Rv_s$, where $v_s$ is the VEV of the singlet $S$. In the dark matter density calculations we will refer to the $\beta_M$-mass generating scenario using Eq.~(\ref{Higgs}) as scenario I and to the one using Eq.~(\ref{singlet}) as scenario II.

Finally note that similar to~\cite{Bajc:2007zf}, gauge invariance allows us to write a Yukawa coupling
\begin{eqnarray}
& & y_w^i H^T(i\tau^2)\omega L_i+y^i_\beta H^T(i\tau^2)\beta L_i + {\rm{c.c.}} \nonumber \\
&=& -\frac{v+h}{\sqrt{2}}[y^i_w(\sqrt{2}w^+e_i+w^0\nu_i)+y^i_\beta\beta\nu_i] + {\rm{c.c.}}
\label{danger_zone}
\end{eqnarray}
However, this coupling is disastrous if we were to have the ``neutralino" as a DM candidate, since the above interaction would allow a decay into light SM leptons. Note that the interaction (\ref{danger_zone}) does not respect the $Z_2$-symmetry discussed above and it can thus be excluded by resorting to a symmetry principle. This symmetry is analogous to the $R$-parity of MSSM and we implement it in the following to guarantee the stability of the DM candidate.

\subsection{Mixing patterns and interactions}
\label{massmix}

Next we specify the mixing pattern between $w^0_M$ and $\beta_M$ states and work out the couplings to the weak gauge bosons in the mass eigenbasis.  Combining the mass terms introduced in the previous section we can write the complete mass Lagrangian for the neutral sector of the theory as follows:
\begin{eqnarray}
{\mathcal{L}}_{\chi} = -\frac{1}{2} 
				\left( \begin{array}{cc} \overline{w}^0_R & \overline{\beta}_R \end{array} \right)
				\left( \begin{array}{cc} M_L   & m_D \\
				                         m_D & M_R     \end{array} \right)
				\left( \begin{array}{cc} w^0_L \\
										 \beta_L    \end{array} \right) + {\rm{h.c.}} 
\label{ecosmo3}
\end{eqnarray}
The mass matrix contains the left handed and right handed Majorana mass terms $M_L$ and $M_R$ from Eqs.~(\ref{doublet2}) and (\ref{Higgs}) and the Dirac mass term $m_D$ from Eq.~(\ref{mixmass2}). Diagonalizing the Lagrangian (\ref{ecosmo3}) we get two mass eigenvalues 
\begin{equation}
\lambda_\pm = \frac{1}{2}( M_L + M_R \pm \sqrt{(M_L - M_R)^2 + 4m_{D}^{2}})\,.
\end{equation}
We are assuming that all masses in Eq.~(\ref{ecosmo3}) are real, but this does not imply that $\lambda_\pm$ are necessarily positive. To ensure positivity we define the physical masses as $m_\pm = \rho_\pm \lambda_\pm$ where the extra phase $\rho_\pm$ is chosen to give $m_\pm \geq 0$ always. Using appropriate global field-redefinitions we can always make $M_L+M_R \ge 0$ which implies that $\rho_+ = +1$ and $m_+ \ge m_-$. We can then identify the heavier neutral state as $\chi_1 \equiv \chi_+$ and the lighter state with $\chi_2 \equiv \chi_-$. (For a more detailed discussion see ref.~\cite{Kainulainen:2009rb}.) Because $|M_L|$ plays also the role of the mass of the Dirac field, we have the additional constraint that $|M_L| > \tilde M$, where $\tilde M$ is the LEP limit for the mass of a charged particle coupling to $Z$-boson with a weak interaction strength. After these identifications we can write the Majorana mass eigenstates as
\begin{eqnarray}
\chi_1 & = &\cos{\theta}(w^{0}_{L} + \rho_1{w^{0}_{R}}) - \sin{\theta}(\rho_1\beta_R + {\beta_{L}}) \nonumber \\
\chi_2 & = &\sin{\theta}(w^{0}_{L} + \rho_2{w^{0}_{R}}) + \cos{\theta}(\rho_2\beta_R + \beta_{L}) \,,
\label{ecosmo4}
\end{eqnarray}
where the mixing angle satisfies $\tan 2\theta = 2m_D/(M_R-M_L)$.    

For the $\chi_2\bar{\chi_2}$ annihilation cross section calculations we need to express the weak currents (\ref{ew_currents}) and the effective Higgs interactions arising from Eqs.~(\ref{doublet2}, \ref{mixmass2}, \ref{Higgs}) and (\ref{singlet}) in the mass eigenbasis. From the Eqs.~(\ref{ew_currents}) we see that the neutral $w^0_M$ couples only to the charged vector bosons $W^{\pm}$ while the singlet $\beta_M$ of course does not have any gauge interactions. To rewrite the charged current in the mass eigenbasis we invert the relations (\ref{ecosmo4}) to give:
\begin{eqnarray}
w^{0}_{L} & = & \cos{\theta}\chi_{1L} + \sin{\theta}\chi_{2L}  
\nonumber \\ 
w^{0}_{R} & = & \cos{\theta}\rho_1\chi_{1R} + \sin{\theta}\rho_2\chi_{2R} 
\label{ecosmo4b}
\end{eqnarray}
Dividing (\ref{ecosmo3}) to chiral components and using (\ref{ecosmo4b}) we now easily find
\begin{eqnarray}
 {\mathcal{L}}_W =  
 g W^{-}_{\mu}\overline{w}_D \Big[ \cos\theta (v_1 - a_1\gamma^5)\chi_{1} 
                                  +\sin\theta (v_2 - a_2\gamma^5)\chi_{2}  \Big] +  {\rm{h.c.}}
\label{ecosmo5}
\end{eqnarray} 
where 
\begin{equation}
v_i \equiv \frac{1}{2}(1 + \rho_i)  \quad {\rm and}Ê\quad
a_i \equiv \frac{1}{2}(1 - \rho_i)\,.
\label{viai}
\end{equation}
From equations (\ref{ecosmo4}) one sees that with small mixing angles the heavier state $\chi_1$ couples strongly with $W$-boson, whereas the lighter state $\chi_2$ is mainly a singlet whose interactions are suppressed by $\sin\theta$. This suppression is the crucial element that will enable us to get the correct dark matter density. Note that the $\chi_2$-interaction involves either a vector current or an axial vector current depending on the sign of the phase factor $\rho_2$. Of course we already know that $\rho_1 = +1$ so that the $\chi_1$-current is always vectorial. However, in what follows we will only need the charged current interactions involving $\chi_2$.

The interactions between the mass eigenstates and the effective composite scalar field $h$ can be directly read off from the lagrangians (\ref{doublet2}, \ref{mixmass2}, \ref{Higgs}) and (\ref{singlet}). They can be converted to the mass eigenbasis by use of (\ref{ecosmo4b}) and 
\begin{eqnarray}
\beta_{L} & = & - \sin{\theta}\chi_{1L} + \cos{\theta}\chi_{2L}  
\nonumber \\ 
\beta_{R} & = & - \sin{\theta}\rho_1\chi_{1R} + \cos{\theta}\rho_2\chi_{2R} \,.
\label{ecosmo4c}
\end{eqnarray}
Both scenarios for the  Majorana mass for the $\beta_M$-field can be combined in a single notation as follows:
\begin{equation}
{\mathcal{L}}_{\chi h}  = -\frac{gm_{2}}{2M_W}
 \Big( C^{h}_{22} h \overline{\chi}_2\chi_2 + \frac{C^{h^2}_{22}}{v} h^2  \overline{\chi}_2\chi_2                           +  C^{h}_{12} h \overline{\chi}_1(v_{12}-a_{12}\gamma_5)\chi_2 \Big) + \cdots, 
\label{ecosmo5b}
\end{equation}
where $v_{12}$ and $a_{12}$ can be found from Eq.~(\ref{ecosmo5b}) replacing $\rho_i \rightarrow \rho_{12} \equiv \rho_1\rho_2$, and we have left out the interactions will not be needed in our calculations in the next section. Moreover, $v$ is the vacuum expectation value of $h$ and the coefficients $C^{h}_{22},C^{h^2}_{22}$ and $C^{h}_{12}$ are listed in table~\ref{Chtable}. We have also left out the interactions between the singlet $s$ and $\chi_2$ in the scenario II by the assumption that $s$-field is very heavy.
\TABLE[t]{
\begin{tabular}{| l | c | c | c |}
\hline
Scenario & $C^{h}_{22}$ & $C^{h^2}_{22}$   & $C^{h}_{12}$ \\ \cline{1-4}
    I    & $1$          & $\frac{1}{2}$    & $0$          \\ \cline{1-4}
    II    &  $\sin^2{\theta}(1+r_{12}\cos^2{\theta})$ 
                        & $\frac{1}{2}C^{h}_{22}$ 
            & $\rho_{12} \sin 2 \theta(1-r_{12}\sin^2{\theta})$ \\ \cline{1-4}
\end{tabular}
\caption{Shown are the coefficients of the different type of composite Higgs - $\chi$ -interactions from the both $\beta$-mass generating scenarios. We defined $\rho_{12} \equiv \rho_1\rho_2$ and $r_{12} \equiv 1 - \rho_{12}\frac{m_1}{m_2}$.}
\label{Chtable}
}

\section{Relic density}
\label{results}

The relic abundance of the lighter stable neutral particle $\Omega_{\chi_2}$ can now be computed in the standard way. First, the number density $n_{\chi_2}$ follows from the usual Lee - Weinberg equation \cite{Lee:1977ua,Kainulainen:2009rb}:
\begin{eqnarray}
\frac{\partial f(x)}{\partial x} 
        = \frac{\langle v\sigma\rangle 
         m_2^3 x^2}{H} (f^2(x)-f_{\rm eq}^2(x)) \,,  
\label{ecosmo1}
\end{eqnarray}
where we have introduced the scaled variables  $f(x) \equiv n_{\chi_2}(x)/{s_E}$  and $x \equiv s_E^{1/3}/m_2$, where $s_E(T)$ is the entropy density at the temperature $T$. Assuming a standard expansion history of the universe the Hubble parameter is $H = (8\pi\rho/3M_{\rm Pl}^2)^{1/2}$, where $\rho(T)$ is the energy density at the temperature $T$. Finally, the average annihilation rate $\langle v\sigma \rangle$ is computed from the expression~\cite{Gondolo:1990dk}:
\begin{equation}
   \langle v \sigma \rangle = 
       \frac{1}{8m_2^{4}TK^{2}_2(\frac{m_2}{T})}
      \int_{4m_2^2}^{\infty}ds
                        \sqrt{s}(s-4m_2^2)K_1(\frac{\sqrt{s}}{T})
                        \sigma_{\rm tot}(s)
\label{ecosmo6}
\end{equation}
where $K_i(y)$'s are modified Bessel functions of the second kind and $s$ is the usual Mandelstam invariant. For the total cross section $\sigma_{\rm tot}$ we included the $\chi_{2}\overline{\chi}_{2}$ annihilation to the $f\bar{f}$, $WW$, $ZZ$ and $hh$ final states, where $f$ is any standard model fermion, $W^{\pm}$ are the charged vector bosons, $Z$ is the neutral vector boson and $h$ is the SM like composite Higgs boson. (The importance of the gauge boson final states for heavy dark matter particles was discovered in ref.~\cite{Enqvist:1988we}.) The matrix elements for these processes can be found in the Appendix. Note that the $Zh$ final state is not open at tree level, because the neutral adjoint states does not couple directly to $Z$; for the same reason both the $f\bar f$ and $WW$-channels are lacking their usual leading s-channel $Z$-exchange diagrams~\cite{Kainulainen:2009rb}. We omitted annihilations to technifermions because these rates would be at best but a small correction to already subleading fermionic channel. We also neglected annihilations to the singlet final states in the scenario II, by assumption that $s$ is a very heavy state. 

\subsection{Results}

Given the averaged cross section, entropy density and the expansion rate of the universe the Lee-Weinberg equation (\ref{ecosmo1}) is easily solved for $f(0)$ giving the ratio of the $\chi_{2}$-number density and entropy density today. The relic density parameter for adjoint Majorana dark matter particle then follows from
\begin{eqnarray}
\label{ecosmo9}
\Omega_{\chi_{2}}\simeq 5.5 \times 10^{11} \, \Big(\frac{m_2}{{\rm TeV}}\Big) f(0)\,,
\end{eqnarray}
where we have used $H_0 \approx 71\,\rm km/sec/Mpc$ for the current expansion rate of the universe.  Our results are shown in Figs.~\ref{omegaresultsrhoN} and
~\ref{omegaresultsrhoN}. Similarly to the case studied in ref.~\cite{Kainulainen:2009rb} we find that $\Omega_{\chi_2}$ is most sensitive to the WIMP mass $m_2$ and to mixing angle $\sin\theta$. It is also strongly sensitive to the mass of the light Higgs field $m_H$. Moreover, the results are very sensitive on the scenario used to create the right chiral Majorana mass, and on the sign of the phase $\rho_{12}$. On the other hand, they depend only very weakly on the charged field and the the heavier neutral field masses. With this in mind we fixed $m_1 = 2m_2$ in all our relic density calculations. This condition also sets, for any given triplet ($m_2,\sin\theta,\rho_{12}$), the mass of $w_D$ through the relation $m_{w_D} \equiv |M_L| = |m_1 \cos^2{\theta} + \rho_{12}m_2 \sin^2{\theta}|$.

\FIGURE[t]{\epsfig{file=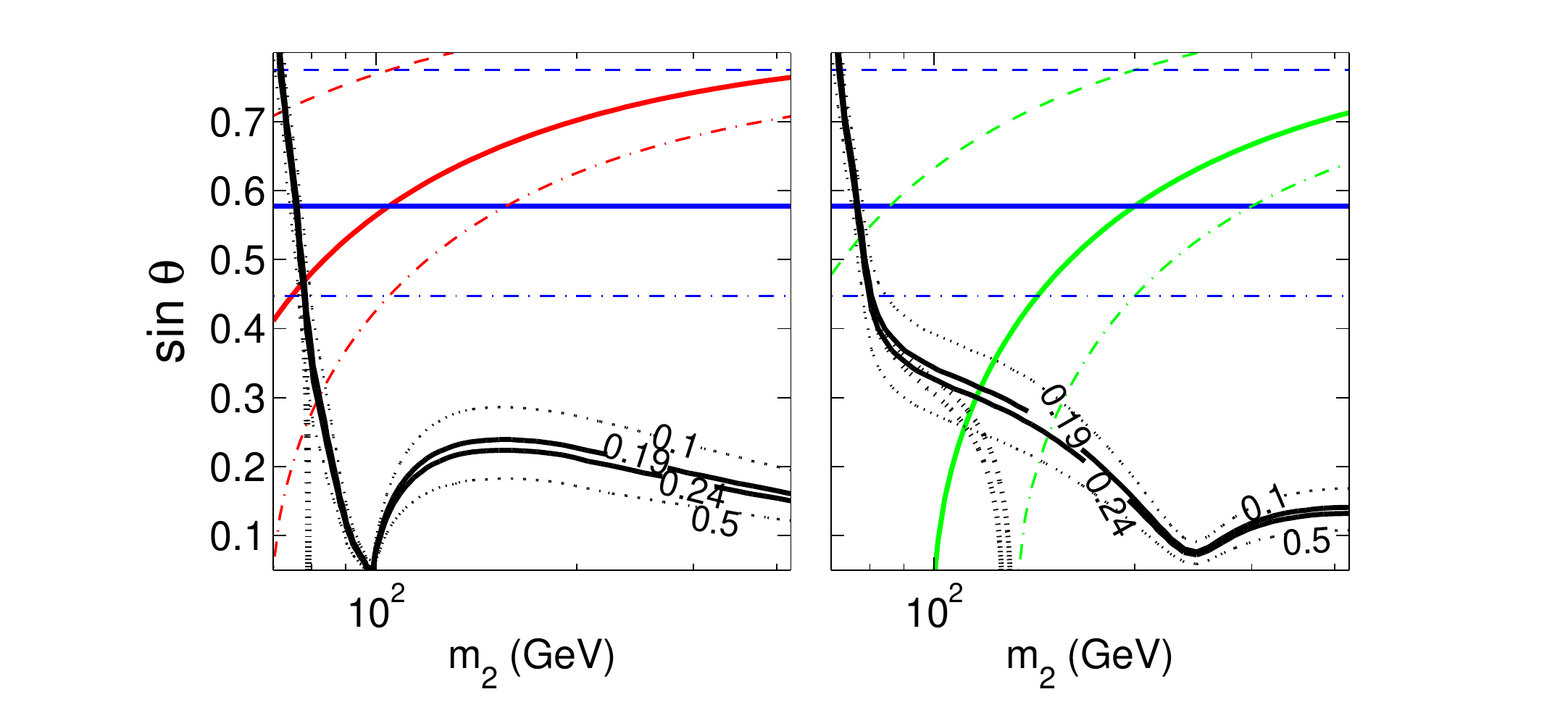,width=14.5cm}
\caption{Shown are the constant $\Omega_{\chi_2}$ contours as a function of WIMP mass and the mixing angle $\sin{\theta}$ with $\rho_{12} = -1$. Scenario I is shown by thick black dotted lines and scenario II by thick solid black and thin dotted lines. In left picture Higgs-boson mass is 200 GeV and in right picture 500 GeV. The area between the thick lines is favoured by the WMAP results for dark matter density. Area to the left from thick red curve in the left panel is excluded by the LEP limit $m_{w_D} < 104.5$ GeV, assuming $m_1/m_2 \equiv A \ge 2$. The dash and dash-dotted lines shown the same limit assuming that $A=4$ and $A=1.5$ respectively. Similar (green) curves on the right panel show these constraints assuming an improved limit $m_{w_D} < 200$ GeV. The (blue) horizontal lines are the upper limits for the $\sin{\theta}$ coming from the requirement that $m_2 < m_{w_D}$, where the thick solid, thin dashed and thin dash-dotted lines again correspond to the cases $A=2$, 4 and 1.5 respectively.}
\label{omegaresultsrhoN}}

In Fig.~\ref{omegaresultsrhoN} we plot constant $\Omega_{\chi_2}$-contours in the $(m_2,\sin\theta)$-plane in the case $\rho_{12} = -1$.
Shown are the $\Omega_{\chi_2}=0.19$ and $\Omega_{\chi_2}=0.24$ contours both in scenario I (thick black dotted lines) and in scenario II (thick black solid lines), which bound the region where the dark matter density is consistent with the WMAP results~\cite{Dunkley:2008ie}. In addition the two thin dotted lines show the results for the scenario II over a larger range of relic densities: $\Omega_{\chi_2}=0.1$ and $\Omega_{\chi_2}=0.5$. In the left panel we used $m_H = 200$ GeV and the right panel $m_H = 500$ GeV. The WIMP mass turns out to be very strongly constrained from above in the scenario I. This result follows from the fact that the Higgs-interactions are not suppressed by the mixing angle in this scenario (see table~\ref{Chtable})), whereby the annihilation cross section becomes very large and suppresses $\Omega_{\chi_2}$ below the required level at large $m_2$. Note that the scenario I is not excluded by observations for these parameters; it simply cannot provide enough dark matter. In the scenario II also the effective Higgs interactions are mixing angle suppressed and consistent solutions are found also for $m_2 > m_H$. The dip in the accepted values for $\sin\theta$ at $m_2 \approx m_H/2$ is caused by the enhanced interaction strength at the Higgs pole. It is evident that values of $m_H$ strongly influence the predictions. In figure~\ref{omegaresultsrhoP} we plot similar results in the case $\rho_{12} = +1$. The results are found to differ from the case $\rho_{12} = -1$ quite significantly. This difference arises both from the strong explicit $\rho_{12}$-dependence of the scalar-couplings listed in the table~\ref{Chtable}, and from the induced $\rho_{12}$-dependence in the matrix element of the $WW$-channel cross section (see Appendix).

\FIGURE[t]{\epsfig{file=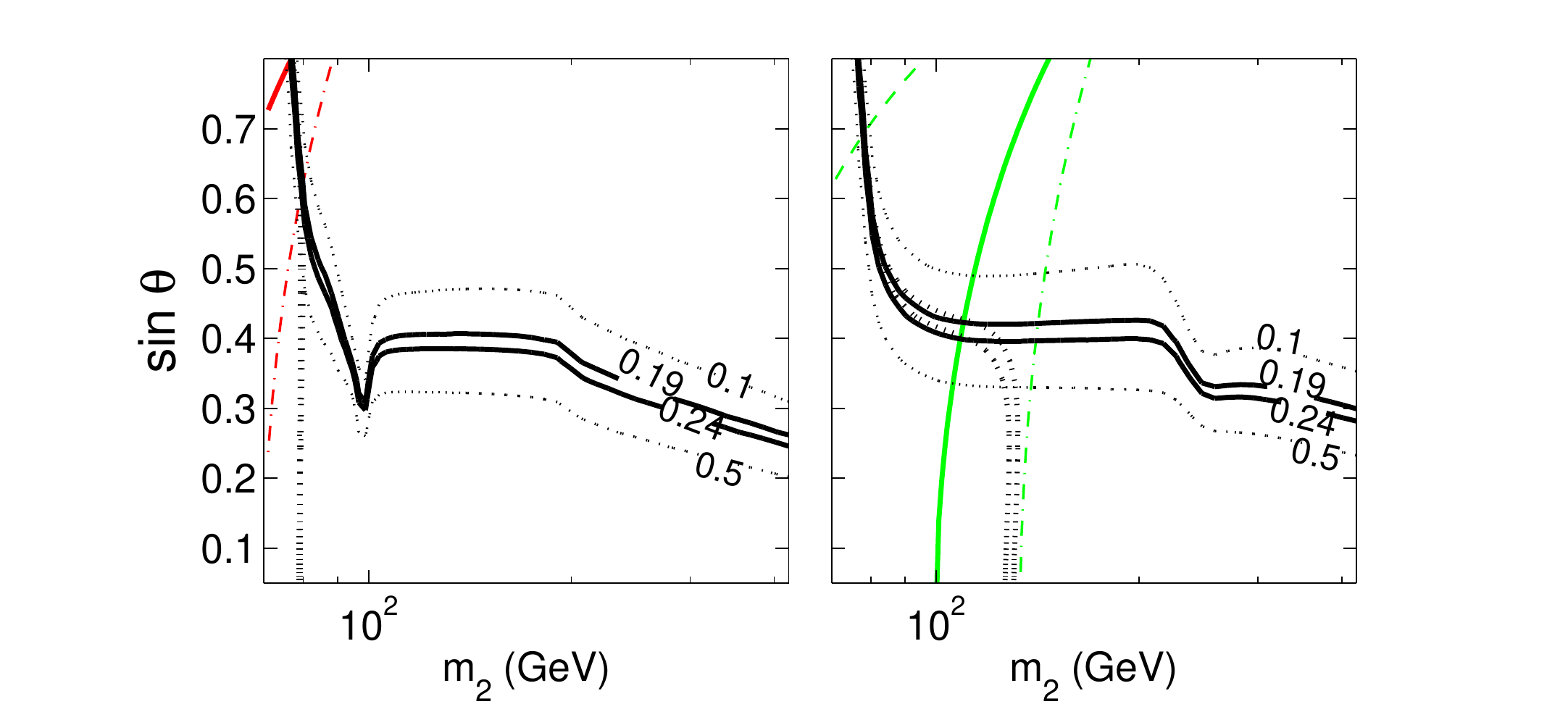,width=14.5cm}
\caption{Same as in Figure \ref{omegaresultsrhoN} but for $\rho_{12} = +1$.}
\label{omegaresultsrhoP}}

Figs.~\ref{omegaresultsrhoN} and \ref{omegaresultsrhoP} display also the current observational constraints on the model parameters. Because $\chi_2$ does not couple to the $Z$-boson, we get no constraints from the current direct dark matter searches however, and due to the same reason $\chi_2$ evades also the constraints coming from $Z$-decay measurements by the LEP experiment. Indeed, currently the only observational constraint to the model comes from the LEP lower bound for the mass of the new {\em charged} state $w_D$: we will use conservatively $m_{w_D} > \tilde{M}$, where $2\tilde M \approx 209$ GeV is the maximum CMS energy of LEP II. As was discussed above, the charged field mass is equal in magnitude to the left-chiral Majorana mass $m_{w_D} = |M_L|$, and hence not an independent parameter in our model. Inverting the relation $m_{w_D} = |m_1 \cos^2{\theta} + \rho_{12}m_2 \sin^2{\theta}| > \tilde M$ gives the conditions:
\begin{equation}
\sin{\theta} < \sqrt{\frac{ A  - \sfrac{\tilde{M}}{m_2}}{A  - \rho_{12} } }
\qquad {\rm or} \qquad
\sin{\theta} > \sqrt{\frac{ A  + \sfrac{\tilde{M}}{m_2}}{A  - \rho_{12} } } \,,
\label{ecosmo9b}
\end{equation}
where we have set $m_1 = A m_2$.The upper bound (first inequality) comes from $M_L > \tilde M$ and the lower bound (second inequality), which defines another disjoint allowed region, from $M_L < -\tilde M$. In addition, we must require that $w_D$ remains heavier than our WIMP: $m_{w_D} > m_2$. Thich implies conditions similar to Eq.~(\ref{ecosmo9b}), where $\tilde M/m_2 \rightarrow 1$. The new lower bound obtained in this way could only be satified by $\sin\theta = 1$ and $\rho_{12}=-1$. Moreover, the regions defined by the new upper bound and the old lower bound do not overlap, and are thus incompatible. The only available solution then is the overlap of the regions satisfying both upper bounds:
\begin{equation}
\sin{\theta} < \sqrt{\frac{ A  - {\rm max}(1,\sfrac{\tilde{M}}{m_2})}{A  - \rho_{12} } }\,.
\label{ecosmo9c}
\end{equation}

The constraints coming from in Eq.~(\ref{ecosmo9b}) are shown  
by the red and green tilted curves in Figures~\ref{omegaresultsrhoN} and~\ref{omegaresultsrhoP} for three different values for the ratio $A =1.5$, 2 and 4. The thick solid lines correspond to the case $A = 2$, which we also used to compute the $\Omega_{\chi_2}$-contours. (Remember that $\Omega_{\chi_2}$ is only very weakly dependent on $A$ however.) The interpretation of these curves is that the area to the left of them is excluded for any $m_1/m_2 \leq A$. The straight horizontal lines in Fig.~\ref{omegaresultsrhoN} correspond to constraint $m_{w_D} > m_2$, computed with the same set of values for $A$. Now the area above the lines is excluded by the stability of $\chi_2$. Note that in the case $\rho_{12}=+1$, shown in Fig.~\ref{omegaresultsrhoP}, $m_{w_D} > m_2$ always and no stability bounds exist. The combined observational and stability bound Eq.~(\ref{ecosmo9c}) is then satisfied to the right and below of the curves described above, for each given value of $A$ respectively. In left panels we have used the actual LEP limit with $\tilde M = 104.5$ GeV, while on the right panels we show how these limits would improve if the bound on $m_{w_D}$ was raised to $\tilde M = 200$ GeV in future experiments, such as LHC. Note that these limits are independent of $m_H$; they were split to different panels just for the sake of clarity. Note also that the exclusion regions in the case $\rho_{12} =-1$ are substantially larger than in the case $\rho_{12} =+1$. This feature follows from the $\rho_{12}$-dependence of $m_{w_D}$ and its effect is explicit in Eq.~(\ref{ecosmo9c}). 

\subsection{Future observational constraints}

The WIMP-$Z$-boson interactions play a crucial role in most dark matter detection modes, including the $Z$-mediated WIMP elastic scatterings off atomic nuclei in the direct dark matter searches and the $Z$-mediated WIMP annihilation to detectable daughter particles in indirect detection modes. Thus the fact that $\chi_2$ has no coupling to $Z$-boson makes it very hard to observe in dark matter searches. Indeed, the dominant interaction channel for $\chi_2$ in direct searches is through the spin-independent Higgs-mediated $t$-channel interaction, which is very strongly suppressed by the small couplings between the Higgs particle and the quarks within the nucleons making up the nuclei in the detector material. Similarly, the dominant indirect detection mode for $\chi_2$ proceeds through an $s$-channel Higgs annihilation, which is also suppressed by the small Higgs couplings to the light SM-fermions.  As a result, the current observational data from DM searches places no significant limits on the regions of model parameters shown in Figs.~\ref{omegaresultsrhoN} and \ref{omegaresultsrhoP}. 

\FIGURE[t]{\epsfig{file=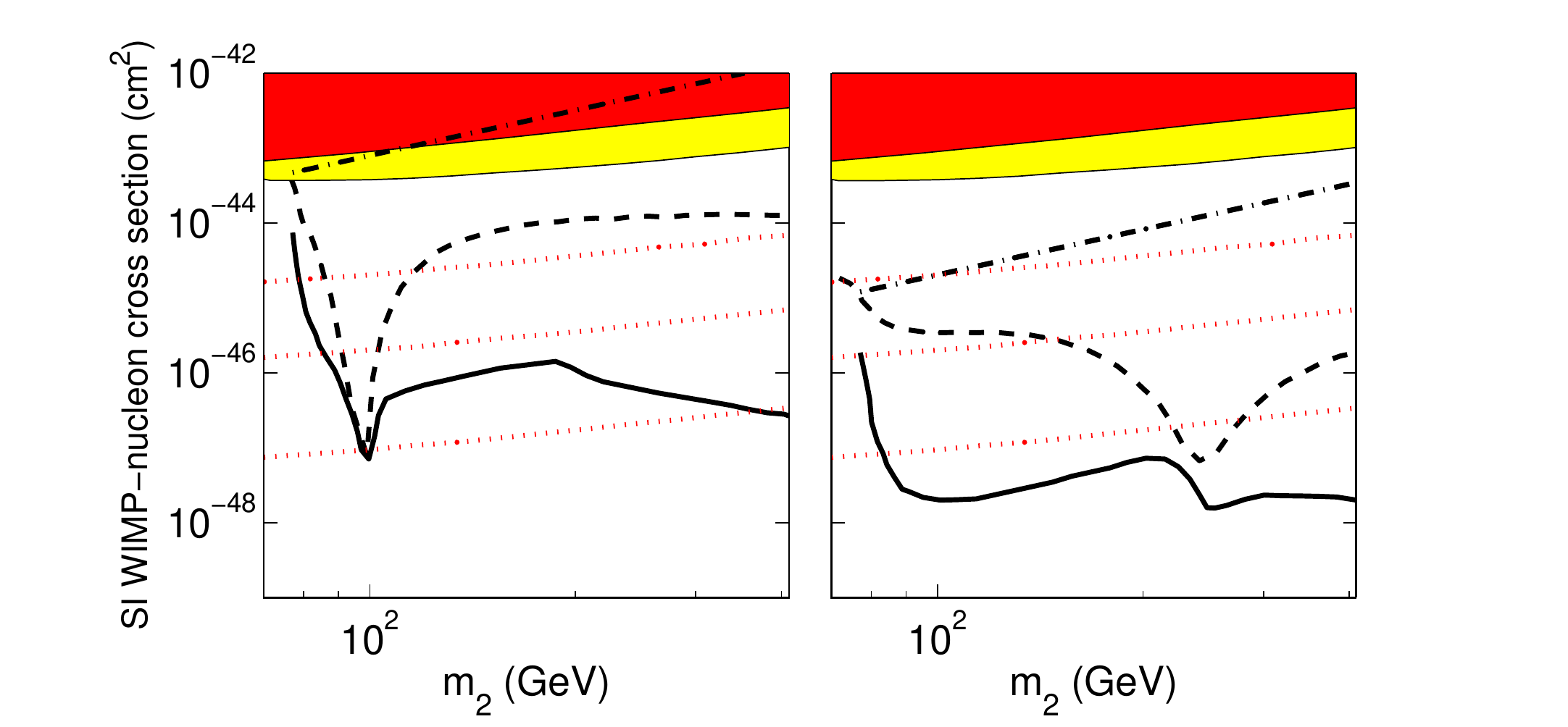,width=14.5cm}
\caption{Shown are the constrains and model predictions for the spin-independent WIMP-nucleon cross section. The yellow area is excluded by the new CDMS II results~\cite{Ahmed:2009zw} and the red area by the XENON10 (2007, Net 136 kg-d) results~\cite{Angle:2007}. In left picture the model predictions were calculated with $m_H=200$GeV and in right picture with $m_H=500$GeV. The black solid curves correspond to $\rho_{12}=+1$ and dashed curves to $\rho_{12}=-1$ in the mass Scenario II. The thick dash-dotted lines correspond to the (SM-like) cross section of the mass scenario I. Finally, the red dotted lines display the projected sensitivity of the XENON100 (up), the upgraded XENON100 (middle) and the XENON1T (low) experiments~\cite{Aprile:2009yh}. We used the tools of ref.~\cite{dmtools} to produce all experimental XENON-limits.}
\label{cryolimits}}

Some DM-detection limits still do exist. In Fig.~\ref{cryolimits} we display the current limits on the spin-independent WIMP-nucleon scattering cross section from the cryogenic dark matter searches in the CDMS II~\cite{Ahmed:2009zw} (light yellow area) and the XENON10~\cite{Angle:2007} (dark red area) experiments. Also shown in the figure are the predictions for the dominant Higgs mediated $\chi_2$-nucleon scattering cross section, which in zero momentum transfer limit reads:
\begin{eqnarray}
\label{ecosmo11}
\sigma_{n}^{0} = (C^{h}_{22})^2 \frac{8 G_F^2}{ \pi}\frac{m_{2}^2 m_{n}^2}{m_{H}^4} K^2 \mu_{n}^{2} \,,
\end{eqnarray}
where $m_{n}$ is nucleon mass, $m_{H}$ is the Higgs boson mass and $K \equiv (1/m_{n}) \sum_{q} \langle n|m_q \bar{q}q|n \rangle \approx 1/3$ is the normalized total scalar quark current in the nucleon. For the individual currents $\langle n|m_q \bar{q}q|n \rangle$ we use the values given in ref.~\cite{Gondolo:1996qw}. $\mu_{n}$ is the WIMP nucleon reduced mass and the scenario-, $\rho_{12}$- and $\sin\theta$-dependent $C^{h}_{22}$-factor is given in Table (\ref{Chtable}). The mixing angle is always chosen to produce the favourable cosmological dark matter density $\Omega_{\chi_2}(m_2,\theta,\rho_{12})=0.214$. This condition leads to a constraint $\theta =\theta(m_2,\rho_{12})$, which through the $C^{h}_{22}$-dependence induces the nontrivial mass dependence of the predictions seen in Fig.~\ref{cryolimits}. The continuous and dashed lines correspond to the predictions of the scenario II with $\rho_{12}=+1$ and $\rho_{12}=-1$ respectively.  Of course, in scenario I the coupling $C^h_{22}$ is $\theta$- and $\rho_{12}$-independent and hence also the predicted $\sigma_n^0$ is independent of the model parameters apart from the Higgs mass. The resulting simple $\sigma^n_0 \sim m_{2}^2$-dependence is shown by the dash-dotted lines in Fig.~\ref{cryolimits}.

As anticipated, the present cryogenic limits place essentially no limits on the scenario II results. In scenario I the CDMS II experiment actually does provide a nontrivial limit for moderately low Higgs masses; for $m_H = 200$ GeV we see that the data just about rules out a cosmologically interesting $\chi_2$. Conversely, for $m_2\approx 80$ GeV this solution is consistent with the tentative dark matter signal reported by the CDMS II experiment~\cite{Ahmed:2009zw}. For larger $m_H$ the cosmologically interesting window extends to about $m_2 \approx 200$ GeV. The situation changes markedly with the upcoming XENON experiments, whose predicted sensitivity on the WIMP-nucleon cross section are shown by the red dotted lines in Fig.~\ref{cryolimits}. In particular the scenario I will either be ruled out or verified as DM, and the XENON1T experiment will be able to put interesting bounds on parameters also on the scenario II, in particular if the Higgs field is again relatively light.

We have also considered indirect limits from detection of WIMP annihilation products in IceCube and Super-Kamiokande experiments. Again the current data places no significant limits on cosmologically interesting model paramters. The projected IceCube results could in principle provide limits comparable to those from XENON100 and its upgrade. However, the precise limits would depend on the estimation on the efficiency of the WIMP-capture in the sun and earth as well as on the precise spectrum of the WIMP annihilation daughter particles. These details are strongly model dependent and we do not try to estimate them here.

Finally, let us briefly discuss the precision electroweak constraints in this model. Most severe constraints in Technicolor model building arise from the $S$ and $T$ parameters. As shown earlier, MWT which also underlies our model here is compatible with electroweak precision data. However, we need to consider possible contributions due to the adjoint matter fields which we have added into the theory. 

Let us first consider $S$-parameter, defined as
\be
S=-\frac{16\pi}{M_Z^2}(\Pi_{3Y}(M_Z^2)-\Pi_{3Y}(0)).
\ee
It follows from Eq. (\ref{ew_currents}) that the SU$_L$(2) adjoint fermions do not couple to the hypercharge at all, and their contribution to $S$ is identically zero. Then consider the $T$ parameter which measures the new physics contributions to the breaking of the custodial isospin symmetry. If the charged and neutral members of the adjoint fermion triplet were degenerate, their contribution to $T$ would be identically zero. This would be the case if the ``bino" was superheavy and decoupled from the low energy theory. However, in the general case we have investigated the neutral and charged mass eigenstates are split, $\delta M\equiv M_{w_D}-M_{w_0}$, and the resulting contribution to $T$ is expected to be nonzero. Roughly, its magnitude is set by the amount of isospin breaking which is of the order of $\delta M/M_Z$.

Therefore, a more complete analysis within this model would be desirable with complete set of precision parameters, $S$,$T$,$Y$,$V$,$W$ (for definitions, see e.g. \cite{Maksymyk:1993zm}) taken into account. Here we just remark that even if the adjoint triplet contributes to $T$-parameter, this can be compensated for by the contribution of the fourth generation leptons present in the MWT model. If one allows for the most general mass and mixing patterns of these leptons, either positive or negative contributions to $T$ can be generated \cite{Antipin:2009ks} and this can compensate for the contribution from the triplet.

\section{Conclusions and outlook}
\label{outlook}

We have considered a new novel dark matter candidate in the context of the minimal walking Technicolor model, originally proposed to avoid the hierarchy problem associated with the fundamental scalar particle in the standard model.
We showed that when the original MWTC model is extended by inclusion of new adjoint SU(3) and SU$(2)_L$ fermions the model can predicts a very good 1-loop unification of the standard model gauge couplings without a need of supersymmetry~\cite{Gudnason:2006mk}. We then considered the case where the adjoint SU$(2)_L$ triplet is allowed to mix with a new singlet state through nonrenormalizable interactions with effective composite scalars. The lighter state in the mixture of the new sterile state and the neutral member of the triplet has suppressed couplings to weak gauge bosons, and its stability can be garanteed by the assumption of a new $Z_2$-symmetry analogous to the R-parity in the MSSM. We showed that this state is a good DM candidate over a range of parameters with $m_2 \gsim 70$ GeV. We also computed the present and expected near future observational bounds on the model. The best current bounds come indirectly from the LEP constraint on the mass of the new charged lepton $w_D$; this state is constructed from the two remaining states the original adjoint Weyl triplet and its mass is therefore nontrivially related to the masses in the neutral sector. Even this bound cannot rule out any cosmologically interesting solutions however. The future direct dark matter searches, such as XENON100 and XENON1T were shown to have be able to rule out significant part of the parameter space. However, even larger experiments would be needed to fully cover the entire parameter space consistent with the observed dark matter abundance in the universe.

In ref.~\cite{Kainulainen:2009rb} a closely related scenario for a dark matter particle with suppressed weak couplings was considered. In that work the DM particle was identified as a mixture of singlet neutrino and a new fourth family doublet neutrino with the usual standard model charges. This new doublet is an integral part of the MWTC model; it was first introduced to cancel the Witten anomaly from the theory, and its contribution to the coupling constant running is essential for the successfull gauge coupling unification. 
The doublet WIMP considered in~\cite{Kainulainen:2009rb} has quite different couplings to the standard model gauge fields than does the adoint WIMP considered in this paper; in particular the adjoint WIMP, in contradiction with the doublet WIMP, does not couple at all to the $Z$-boson. This is the single most important reason why the observational constraints on the adjoint WIMP are much weaker than those on the doublet WIMP.  The adjoint and doublet dark matter scenarios of the MWTC model, discussed here and in~\cite{Kainulainen:2009rb} can naturally be combined into a unified model where the WIMP is identified as the lightest member of the mixture of the three new neutral particles: the neutral adjoint, the neutral doublet and the new singlet state. This neutral sector is in many ways similar to the neutralino sector in the MSSM and its phenomenology will be discussed elsewhere~\cite{KTV4}.

Let us finally comment on the two possible WIMP signals recently reported by the CDMS II collaboration~\cite{Ahmed:2009zw}. It is perhaps still too early to decide if these events are real, but if they were, they could be accounted for in our model in the mass scenario I with an adjoint WIMP mass $m_2 \approx 80$ GeV and with a light composite Higgs mass $m_H \approx 200$ GeV. Moreover, the signal can easily be made consistent with the doublet dark matter case, and of course also with the more general doublet-adjoint-singlet mixing case discussed above. Also, if the signals persist, they should be very clear in the data from the upcoming XENON experiments, and that obviously would impose very interesting limits on the parameters of our models.

\section*{Acknowledgments}
\noindent JV thanks the Finnish Academy of Sciences and Letters, Vilho, Yrj\"o and Kalle V\"ais\"al\"a foundation, and the Academy of Finland, Graduate School in Particle and Nuclear Physics (Graspanp) for grants.

\section{Appendix}

Here we give some of the necessary building blocks for the construction of the total dark matter annihilation cross sections. The cross sections to final states $f\bar{f}$ and $ZZ$ are simple enough to be given here explicitly. However, the cross sections for the $W^+W^-$ and $HH$ final states are too lengthy and so we will only write down the matrix elements for them. These expressions are still useful because computing the matrix elements for processes involving Majorana fermions is much more tedious than the corresponding quantities with Dirac fields. In particular the role of the relative phases in the case of mixing fields can be problematic~\cite{Kainulainen:2009rb}. It is also easy to compute the cross sections from the matrix elements using algebraic programs such as FEYNCALC. 

Because $\chi_{2}$-field does not couple to $Z$-boson, the annihilation $\chi_{2}\overline{\chi}_{2}  \rightarrow f \overline{f}$ proceeds only through the $s$-channel Higgs exchange diagram at the tree level. The final result for the cross-section is
\begin{equation}  
\sigma_{f \overline{f}}  = \frac{G_{F}^2 \beta_f}{2 \pi s \beta_2} N^{f}_{c} |D_H|^2 (C_{22}^{h})^2 
m_2^2m_f^2 (s-4m_{2}^2)(s-4m_{f}^2).
\label{effmatrixel}
\end{equation}
where, 
\begin{equation}
\beta_X \equiv \Big(1-\frac{4m_X^2}{s}\Big)^{1/2} 
\quad {\rm and}Ê\quad 
D_X \equiv \frac{1}{s - m_X^2 + i \Gamma_X m_X}
\label{DH}
\end{equation}
and $N^f_c$ is the color factor of the fermion $f$ (1 for leptons and 3 for quarks), and the mass scenario dependent factor $C^h_{22}$ is given in table \ref{Chtable}. 
The annihilation $\chi_{2}\overline{\chi}_{2} \rightarrow Z Z$ also proceeds only through an $s$-channel Higgs boson exchange. We find:
\begin{equation}    
\sigma_{Z Z}  = \frac{G_{F}^2 \beta_Z}{2 \pi s \beta_2}  |D_H|^2 
(C_{22}^{h})^2 m_2^2  
(s-4m_{2}^2)(s^2-4sm_Z^2+12m_Z^4) \,. 
\label{eZZmatrixel}
\end{equation}
%

The reaction $\chi_{2}\overline{\chi}_{2}\rightarrow WW$ is mediated by the $t$- and $u$-channel charged $w_D$ exchanges and an $s$-channel Higgs boson exchange. The matrix element for this process is
\begin{equation}
{\mathcal M}_{W^{+}W^{-}} = i \rho_2 \frac{g^2}{2} \epsilon_{\mu_1}^{+*}(k_1) \epsilon_{\mu_2}^{-*}(k_2)
                                \bar{v}_{\chi_{2}}(p_1) \Gamma^{\mu_1\mu_2}_{W^{+}W^{-}} u_{\chi_{2}}(p_2), 
\label{eWWmatrixel}
\end{equation}
where $\epsilon^\pm_{\mu}$ are the charged gauge boson polarization vectors and  
\begin{eqnarray}
\Gamma^{\mu_1\mu_2}_{W^{+}W^{-}} & = & 2  \sin^2{\theta} \left[ D_{t}^{w_D} \gamma^{\mu_1} ( \ksl_1 - \psl_1 + \rho_{12} m_{w_D} )  \gamma^{\mu_2} + D_{u}^{w_D} \gamma^{\mu_2}  ( \ksl_2 - \psl_1 + \rho_{12} m_{w_D} ) \gamma^{\mu_1} \right] 
\nonumber \\& -  &  2 m_2 C_{22}^{h}D_H g^{\mu_1 \mu_2} \,, 
\label{eWWmatrixel2}
\end{eqnarray}
where $D_H$ is again given by Eq.~(\ref{DH}) and the $t$- and $u$-channel propagators are 
\begin{equation}
D^{w_D}_a \equiv \frac{1}{a - m_{w_D}^2} \,,
\label{eq:simpleprop}
\end{equation}
where $m_{w_D} = |M_L| =  |m_1 \cos^2{\theta} + \rho_{12}m_2 \sin^2{\theta}|$.  Finally, the annihilation to final state Higgs bosons
$\chi_{2}\overline{\chi}_{2}  \rightarrow H \overline{H}$ proceeds through both $\chi_{1}$- and $\chi_{2}$-mediated $t$- and $u$-channel diagrams as well as an $s$-channel Higgs exchange and the $hh\chi_{2}\chi_{2}$-contact interaction.
The matrix element for this process is
\begin{equation}
{\mathcal M}_{H H} = i \rho_2 \frac{g^2}{2} \frac{m_{2}^{2}}{m_{W}^{2}} \bar{v}_{\chi_{2}}(p_1) A_{H H} u_{\chi_{2}}(p_2).    
\label{eHHmatrixel}
\end{equation}
where
\begin{eqnarray}
A_{H H} & = & 2 (C_{22}^{h})^2 \left[ D_u^{\chi_2}(2m_2- \ksl_1)  + D_{t}^{\chi_2}(2m_2- \ksl_2)\right]  \\  \nonumber 
        & + & \sfrac{1}{2} (C_{12}^{h})^2 \left[ D_{u}^{\chi_1}(m_2 +\rho_{12}m_1 - \ksl_1)  + D_{t}^{\chi_1}(m_2 +\rho_{12}m_1 -\ksl_2)\right] \\
        & + & 3D_H \frac{m_{H}^{2}}{m_2} C_{22}^{h} + \frac{2}{m_2}C_{22}^{h^2}.   \nonumber  
\label{eHHmatrixel2}
\end{eqnarray}
Here the propagators $D_a^{\chi_i}$ are given by expressions analogous to (\ref{eq:simpleprop}) and the factors $C_{22}^{h}$, $C_{12}^{h}$ and $C_{22}^{h^2}$ are again given in the table~\ref{Chtable}.

\end{document}